\documentclass{article}

\usepackage[frozencache,cachedir=.]{minted}
\usepackage{tabu}

\usepackage{microtype}
\usepackage{graphicx}
\usepackage{booktabs} % for professional tables
\usepackage{amsmath}
\usepackage{amsfonts}
\usepackage{mathtools}
\usepackage{caption}
\usepackage{subcaption}
\DeclareMathOperator*{\argmaxA}{arg\,max}

\usepackage{xcolor}
\usepackage{gensymb}
\newcommand{\defcommenter}[2]{%
  \expandafter\newcommand\csname #1\endcsname[1]{%
  {\color{#2}[#1: ##1]}%
  }%
}
\defcommenter{TODO}{red}
\defcommenter{JENNY}{green}
\usepackage{hyperref}
\usepackage[accepted]{mlsys2020}

\mlsystitlerunning{AutoPhase: Juggling HLS Phase Orderings in Random Forests with Deep Reinforcement Learning}

\begin{document}
% \abovedisplayskip=0pt
% \abovedisplayshortskip=0pt
% \belowdisplayskip=0pt
% \belowdisplayshortskip=0pt
% \abovecaptionskip=0pt
% \belowcaptionskip=0pt

%\title{Juggling HLS Phase Orderings in Random Forests with Deep Reinforcement Learning}
\twocolumn[
\mlsystitle{AutoPhase: Juggling HLS Phase Orderings in Random Forests with Deep Reinforcement Learning}
\mlsyssetsymbol{equal}{*}

\begin{mlsysauthorlist}
\mlsysauthor{Qijing Huang}{equal,ucb}
\mlsysauthor{Ameer Haj-Ali}{equal,ucb}
\mlsysauthor{William Moses}{mit}
\mlsysauthor{John Xiang}{ucb}
\mlsysauthor{Krste Asanovic}{ucb}
\mlsysauthor{John Wawrzynek}{ucb}
\mlsysauthor{Ion Stoica}{ucb}
\end{mlsysauthorlist}

\mlsysaffiliation{ucb}{University of California, Berkeley}
\mlsysaffiliation{mit}{Massachusetts Institute of Technology}
\mlsyscorrespondingauthor{Qijing Huang}{qijing.huang@berkeley.edu}
\mlsyscorrespondingauthor{Ameer Haj-Ali}{ameerh@berkeley.edu}

\mlsyskeywords{Compiler, Phase Ordering, Reinforcement Learning}
% unless you've provided the [accepted] option to the sysml2019


% \sysmlsetsymbol{equal}{*}
% \author{Qijing Huang*$^\dag$, Ameer Haj-Ali*$^\dag$, William Moses$^\dag$, John Xiang, Ion Stoica, Krste Asanovic, \\ John Wawrzynek \\
% \{qijing.huang, ameerh, johnxiang, istoica, krste, johnw\}@berkeley.edu, wmoses@mit.edu
% }
% \thanks{\hspace{-0.27cm}\rule{3.15in}{0.4pt}}
%\thanks{\fontsize{9.5}{13.5}\selectfont *Equal contribution.}
%\maketitle
% \thanks{\fontsize{9.5}{13.5}\selectfont \dag  Primary contributor.}}

\begin{abstract}
The performance of the code a compiler generates depends on the order in which it applies the optimization passes.  
Choosing a good order--often referred to as the {\em phase-ordering} problem, is an NP-hard problem. As a result, existing solutions rely on a variety of heuristics.
In this paper, we evaluate a new technique to address the phase-ordering problem: deep reinforcement learning.
To this end, we implement AutoPhase\footnote[3]{}: a framework that takes a program and uses deep reinforcement learning to find a sequence of compilation passes that minimizes its execution time. 
Without loss of generality, we construct this framework in the context of the LLVM compiler toolchain and target high-level synthesis programs. 
We use random forests to quantify the correlation between the effectiveness of a given pass and the program's features. This helps us reduce the search space by avoiding phase orderings that are unlikely to improve the performance of a given program. 
We compare the performance of AutoPhase to state-of-the-art algorithms that address the phase-ordering problem.
In our evaluation, we show that AutoPhase improves circuit performance by 28\%
when compared to using the -O3 compiler flag, and achieves competitive results compared to the state-of-the-art solutions, while requiring fewer samples.
Furthermore, unlike existing state-of-the-art solutions, our deep reinforcement learning solution shows promising result in generalizing to real benchmarks and 12,874 different randomly generated programs, after training on a hundred randomly generated programs.

% for less than ten minutes.

%
% Existing methods for solving the phase-ordering problem are based on trials on per-program basis. 
% Our method uses RL to learn the heuristics from salient program features and can applied pre-model to new programs. 
% We analyze the correlation between the features (applied passes, program features) and next action that optimizes overall rewards by training a random forest.  
% We generate thousands of HLS-synthesizable random programs, train them and run inference/transfer learning on real HLS design. Results shows better than O3 performance on real benchmarks

\end{abstract}
%\IEEEpeerreviewmaketitle
]

\printAffiliationsAndNotice{\mlsysEqualContribution}
%\maketitle

\section{Introduction}
High-Level Synthesis (HLS) automates the process of creating digital hardware circuits from algorithms written in high-level languages.  
Modern HLS tools~\cite{xilinx_vivado_hls,intel_hls,canis2013legup} use the same front-end as the traditional software compilers.  
They rely on traditional software compiler techniques to optimize the input program's intermediate representation (IR) and produce circuits in the form of RTL code. % from the IRs. 
Thus, the quality of compiler front-end optimizations directly impacts the performance of HLS-generated circuit. 

Program optimization is a notoriously difficult task. 
A program must be just in "the right form" for a compiler to recognize the optimization opportunities. This is a task a programmer might be able to perform easily, but is often difficult for a compiler.
%To add to this complexity, often the optimization is hardware-dependent.
Despite a decade of research on developing sophisticated optimization algorithms, 
there is still a performance gap between the HLS generated code and the hand-optimized one produced by experts. %, though it incurs huge costs both in terms of time and human capital. 
% The goal of this paper is to work towards an intelligent HLS frontend that automatically decides which  optimizations and in which order to apply. 
%The main task of a compiler is to take a program written in a high level language and produce optimized executable code to run on specific platforms. This is a notoriously difficult task. 
%To guarantee they produce correct code, compilers make conservative decisions, which can lead to missing optimizations. \JENNY{This is not the problem we try to solve as we didn't modify the passes to be more intrusive}
%Thus, despite decades of research, an expert programmer can still produce code that outperforms a compiler. Unfortunately, manually optimizing code incurs huge costs both in terms of time and human capital. 

%Compilers have been successful in accomplishing two distinct goals: they have automated many expert-programmer techniques and have allowed for programmers to write high-level and algorithmically elegant programs. However, the benefits that compilers provide do not come without cost. Programs must be in just the right form for an optimization pass to recognize it -- a task which a programmer might be able to easily do, but is often difficult for the compiler. Rather than risk producing incorrect code, the compiler must act conservatively, leading to many missed optimizations. Thus, a program manually optimized by a programmer may often outperform a program optimized by a compiler, though at a significant cost both in time and intellectual capital.

In this paper, we build off the LLVM compiler~\cite{lattner2004llvm}. However, our techniques, can be broadly applicable to any compiler that uses a series of optimization passes.
In this case, the optimization of an HLS program consists of applying a sequence of analysis and optimization phases, where each phase in this sequence consumes the output of the previous phase, and generates a modified version of the program for the next phase. Unfortunately, these phases are not commutative which makes the order in which these phases are applied critical to the performance of the output. 
%, each of them A compiler is composed of analyses and optimizations which are simply transformations that modify programs. Unfortunately, since optimizations do not generally commute, it is crucial to apply optimizations in the proper order. 

Consider the program in Figure~\ref{fig:norm1}, which normalizes a vector. Without any optimizations, the \verb|norm| function will take $\Theta(n^2)$ to normalize a vector. However, a smart compiler will implement the \textit{loop invariant code motion (LICM)}~\cite{Muchnick97} optimization, which allows it to move the call to \verb|mag| above the loop, resulting in the code on the left column in Figure~\ref{fig:norm2}. This optimization brings the runtime down to $\Theta(n)$---a big speedup improvement. Another optimization the compiler could perform is \textit{(function) inlining}~\cite{Muchnick97}. With inlining, a call to a function is simply replaced with the body of the function, reducing the overhead of the function call. Applying inlining to the code 
%in Figure \ref{fig:norm2}, 
will result in the code in the right column of Figure~\ref{fig:norm2}.
\begin{figure}[H]
     \centering
         \centering
\begin{minted}[fontsize=\footnotesize]{c}
__attribute__((const))
double mag(int n, const double *A) {
    double sum = 0;
    for(int i=0; i<n; i++){
        sum += A[i] * A[i];
    }
    return sqrt(sum);
}
void norm(int n, double *restrict out,
          const double *restrict in) {
    for(int i=0; i<n; i++) {
        out[i] = in[i] / mag(n, in);
    }
}
\end{minted}
\caption{A simple program to normalize a vector. }
    \label{fig:norm1}
\end{figure}
\begin{figure*}[h]
     \centering
     \begin{subfigure}[b]{0.46\textwidth}
         \centering
\begin{minted}[fontsize=\footnotesize]{c}
void norm(int n, double *restrict out,
          const double *restrict in) {
    double precompute = mag(n, in);
    for(int i=0; i<n; i++) {
        out[i] = in[i] / precompute;
    }
}
\end{minted}
    %\label{fig:norm2}
\end{subfigure}
     \hfill
     \begin{subfigure}[b]{0.46\textwidth}
         \centering
\begin{minted}[fontsize=\footnotesize]{c}
void norm(int n, double *restrict out,
          const double *restrict in) {
    double precompute, sum = 0;
    for(int i=0; i<n; i++){
        sum += A[i] * A[i];
    }
    precompute = sqrt(sum);
    for(int i=0; i<n; i++) {
        out[i] = in[i] / precompute;
    }
}
\end{minted}
     \end{subfigure}
     \caption{Progressively applying LICM (left) then inlining (right) to the code in Figure~\ref{fig:norm1}.}
    \label{fig:norm2}
\end{figure*}
\begin{figure*}[!h]
     \centering
     \begin{subfigure}[b]{0.46\textwidth}
         \centering
\begin{minted}[fontsize=\footnotesize]{c}
void norm(int n, double *restrict out,
          const double *restrict in) {
    for(int i=0; i<n; i++) {
        double sum = 0;
        for(int j=0; j<n; j++){
            sum += A[j] * A[j];
        }
        out[i] = in[i] / sqrt(sum);
    }
}
\end{minted}
    %\label{fig:norm4}
\end{subfigure}
     \hfill
     \begin{subfigure}[b]{0.46\textwidth}
         \centering
\begin{minted}[fontsize=\footnotesize]{c}
void norm(int n, double *restrict out,
          const double *restrict in) {
    double sum;
    for(int i=0; i<n; i++) {
        sum = 0;
        for(int j=0; j<n; j++){
            sum += A[j] * A[j];
        }
        out[i] = in[i] / sqrt(sum);
    }
}
\end{minted}
     \end{subfigure}
          \vspace{-0.3cm}
     \caption{Progressively applying inlining (left) then LICM (right) to the code in Figure~\ref{fig:norm1}.}
     \label{fig:norm3}
     %\vspace{-0.5cm}
\end{figure*}

Now, consider applying these optimization passes in the opposite order: first inlining then LICM. After inlining, we get the code on the left of Figure~\ref{fig:norm3}. Once again we get a modest speedup, having eliminated $n$ function calls, though our runtime is still $\Theta(n^2)$. If the compiler afterwards attempted to apply LICM, we would find the code on the right of Figure~\ref{fig:norm3}. LICM was able to successfully move the allocation of sum outside the loop. However, it was unable to move the instruction setting \verb|sum=0| outside the loop, as doing so would mean that all iterations excluding the first one would end up with a garbage value for sum. Thus, the internal loop will not be moved out.

As this simple example illustrates, the order in which the optimization phases are applied can be the difference between the program running in $\Theta(n^2)$ versus $\Theta(n)$. It is thus crucial to determine the optimal phase ordering to maximize the circuit speeds. Unfortunately, not only is this a difficult task, but the optimal phase ordering may vary from program to program. Furthermore, it turns out that finding the optimal sequence of optimization phases is an NP-hard problem, and exhaustively evaluating all possible sequences is infeasible in practice. In this work, for example, the search space extends to more than $2^{247}$ phase orderings.

The goal of this paper is to provide a mechanism for automatically determining good phase orderings for HLS programs to optimize for the circuit speed. To this end, we aim to leverage recent advancements in deep reinforcement learning (RL)~\cite{sutton1998,hajali2019deep} to address the phase ordering problem. 
%targeting hardware designs.
%Recent advancements in deep reinforcement learning (RL)~\cite{sutton1998} offer opportunities to address the phase ordering challenge. 
With RL, a software agent continuously interacts with the environment by taking actions. Each action can change the state of the environment and generate a "reward". The goal of RL is to learn a policy---that is, a mapping between the observed states of the environment and a set of actions---to maximize the cumulative reward. An RL algorithm that uses a deep neural network to approximate the policy is referred to as a deep RL algorithm. In our case, the observation from the environment could be the program and/or the optimization passes applied so far. The action is the optimization pass to apply next, and the reward is the improvement in the circuit performance after applying this pass. 
%Generally, deep RL algorithms interact with the environment and are rewarded based on their actions. 
%Based on these rewards and environment observations, statistical machine learning is applied to estimate the long-term benefit of the actions and optimize these actions accordingly. 
%In this work, we explore the benefits of applying deep RL algorithms to improve the quality of results (QoR) of HLS.
The particular framing of the problem as an RL problem has a significant impact on the solution's effectiveness. Significant challenges exist in understanding how to formulate the phase ordering optimization problem in an RL framework. 

In this paper, we consider three approaches to represent the environment's state. The first approach is to directly use salient features from the program. The second approach is to derive the features from the sequence of optimizations we applied while ignoring the program's features. The third approach combines the first two approaches. We evaluate these approaches by implementing a framework that takes a group of programs as input and quickly finds a phase ordering that competes with state-of-the-art solutions.  %Furthermore, an agent trained on random programs improves the test random programs by 5\% and 3\% on the nine real benchmarks. 
Our main contributions are:
%\vspace{-0.2cm}
\begin{itemize}
\itemsep 0em 

    %\item Modified the LLVM compiler to extract 
    %program features that describe the program structure.  %describe the number of different operations from a specific HLS program.
    \item Extend a previous work~\cite{huang2019autophase} and leverage deep RL to address the phase-ordering problem.
    \item An importance analysis on the features using random forests to significantly reduce the state and action spaces.
    \item AutoPhase: a framework that integrates the current HLS compiler infrastructure with the deep RL algorithms.
    \item We show that AutoPhase gets a 28\% improvement over -O3 for nine real benchmarks.  Unlike all state-of-the-art approaches, deep RL demonstrates the potential to generalize to thousands of different programs after training on a hundred programs.
\end{itemize}
\vspace{-0.3cm}
\section{background} % 1.5 pages
\label{sec:bg}
\subsection{Compiler Phase-ordering}
Compilers execute optimization passes to transform programs into more efficient forms to run on various hardware targets.
Groups of optimizations are often packaged into ``optimization levels'' , such as -O0 and -O3, for ease. 
%-O0 (no optimization), -O1 (some optimization) -O2 (more optimization), -O3 (most optimization)
While these optimization levels offer a simple set of choices for developers, they are handpicked by the compiler-designers and often most benefit certain groups of benchmark programs. 
The compiler community has attempted to address the issue by selecting a particular set of compiler optimizations on a per-program or per-target basis for software~\cite{triantafyllis2003compiler,almagor2004finding,pan2006fast, ansel2014opentuner}.

Since the search space of phase-ordering is too large for an exhaustive search, many heuristics have been proposed to explore the space by using machine learning. 
Huang~\textit{et al.} tried to address this challenge for HLS applications by using modified greedy algorithms~\cite{huang2013effect,huang2015effect}. It achieved 16\% improvement vs -O3 on the CHstone benchmarks~\cite{hara2008chstone}, which we used in this paper.  %to apply a new optimization to an existing sequence and keep the best one or three sequences at each 
In~\cite{agakov2006using} both independent and Markov models were applied to automatically target an optimized search space for iterative methods to improve the search results.
In~\cite{2003Stephenson}, genetic algorithms were used to tune heuristic priority functions for three compiler optimization passes. 
Milepost GCC~\cite{fursin2011milepost} used machine learning to determine the set of passes to apply to a given program, based on a static analysis of its features. It achieved an 11\% execution time improvement over -O3, for the ARC reconfigurable processor on the MiBench program suite1.
In~\cite{2012Kulkarni} the challenge was formulated as a Markov process and supervised learning was used to predict the next optimization, based on the current program state.
OpenTuner~\cite{ansel2014opentuner} autotunes a program using an AUC-Bandit-meta-technique-directed ensemble selection of algorithms. Its current mechanism for selecting the compiler optimization passes does not consider the order or support repeated optimizations. 
Wang \textit{et al.}~\cite{wang2018}, provided a survey for using machine learning in compiler optimization where they also described that using program features might be helpful. NeuroVectorizer~\cite{haj2019neurovectorizer,hajlearning} used deep RL for automatically tuning compiler pragmas such as vectorization and interleaving factors. NeuroVectorizer achieves 97\% of the oracle performance (brute-force search) on a wide range of benchmarks.

\subsection{Reinforcement Learning Algorithms}
Reinforcement learning (RL) is a machine learning approach in which an agent continually interacts with the environment~\cite{kaelbling1996reinforcement}. In particular, the agent observes the state of the environment, and based on this observation takes an action. The goal of the RL agent is then to compute a policy--a mapping between the environment states and actions--that maximizes a long term reward. 
%by taking actions. These actions are generally the output of a policy that takes as an input a state/observation. The objective of RL is to find a policy that maximizes the reward an agent receives while interacting with the environment.

RL can be viewed as a stochastic optimization solution for solving Markov Decision Processes (MDPs)~\cite{bellman1957}, when the MDP is not known. An MDP 
%models the system with an agent that wants to optimize a given objective (such as the score in a game) by making a sequence of decisions. Given the current state, a decision is made, which leads to a new state. More formally, the Markov decision process 
is defined by a tuple with four elements:
${S,A,P(s,a),r(s,a)}$ where $S$ is the set of states of the environment, $A$ describes the set of actions or transitions between states, $s' {\raise.17ex\hbox{$\scriptstyle\mathtt{\sim}$}} P(s,a)$ describes the probability distribution of next states given the current state and action and $r(s,a):S \times A \rightarrow R$ is the reward of taking action $a$ in state $s$. Given an MDP, the goal of the agent is to gain the largest possible aggregate reward. The objective of an RL algorithm associated with an MDP is to find a decision policy $\pi^*(a|s):s\rightarrow A$ that achieves this goal for that MDP:
 %, by mapping states to actions, that maximize the reward:
\begin{multline}
\label{eq:MDP}
    \pi^* = \argmaxA_\pi \mathbb{E}_{\tau{\raise.05ex\hbox{$\scriptstyle\mathtt{\sim}$}}\pi(\tau)} \left[\sum_{t}^{}r(s_t,a_t) \right] = \\
    \argmaxA_\pi \sum_{t=1}^{T}\mathbb{E}_{(s_t,a_t){\raise.05ex\hbox{$\scriptstyle\mathtt{\sim}$}}\pi(s_t,a_t)}\left[r(s_t,a_t)\right].
\end{multline}

Deep RL leverages a neural network to learn the policy (and sometimes the reward function).
%Recently, Deep RL has achieved impressive results, such as learning to play 49 Atari games with human-level capabilities~\cite{mnih2015}, and defeating the Go world champion~\cite{silver2016}. 
Policy Gradient (PG)~\cite{sutton2000}, %and Deep Q-Network (DQN), commonly referred to as Q-Learning~\cite{watkins1992q}. We chose these algorithms as they are simple, mature, and adequate to solve the problem. 
for example, updates the policy directly by differentiating the aggregate reward $\mathbb{E}$ in Equation~\ref{eq:MDP}:
\begin{multline}
\label{eq:policygradient}
    \nabla_\theta J =  \\
    %\nabla_\theta \mathbb{E}_{\tau{\raise.05ex\hbox{$\scriptstyle\mathtt{\sim}$}}\rho_{\pi(\tau)}} \left[\sum_{t}^{}r(s_t,a_t) \right] = \\
    %\mathbb{E}_{\tau{\raise.05ex\hbox{$\scriptstyle\mathtt{\sim}$}}\rho_{\pi(\tau)}} \left[(\sum_{t}^{}\nabla_\theta log\pi_\theta(a_t|s_t))(\sum_{t}^{}r(s_t,a_t))\right] \approx \\
    \frac{1}{N}\sum_{i=1}^{N} \left[ (\sum_{t}^{}\nabla_\theta log\pi_\theta(a_{i,t}|s_{i,t}))(\sum_{t}^{}r(s_{i,t},a_{i,t})) \right]
\end{multline}
and updating the network parameters (weights) in the direction of the gradient:
\begin{equation}
    \theta \leftarrow \theta+\alpha\nabla_\theta J,
\end{equation}
%where $\alpha$ represents the learning rate.
Note that PG is an on-policy method in that it uses decisions made directly by the current policy to compute the new policy.

Over the past couple of years, a plethora of new deep RL techniques have been proposed~\cite{mnih2016asynchronous,ross2011}. In this paper, we mainly focus on Proximal Policy Optimization (PPO)~\cite{schulman2017proximal}, 
Asynchronous Advantage Actor-critic (A3C)~\cite{mnih2016asynchronous}.
%In contrast, DQN is an off-policy method which takes partially random actions to explore the state space. The DQN's goal is to find which actions will maximize future rewards from a given state. To do so DQN computes a $Q$-function, $Q(s_i, a_i)$ that predicts the future overall reward of taking action $a_i$ from state $s_i$. To compute this $Q$-function, DQN uses a neural network paramterized by weights $\phi$). More formally:

% \begin{equation}
% \label{eq:Dqn}
%     y_i = r(s_i,a_i) + \argmaxA_{a'_i}Q_\phi(s'_i,a'_i)\\
% \end{equation}
% \begin{equation}
%     \phi \leftarrow \argminA_{\phi}\sum_i||Q_\phi(s_i,a_i)-y_i||^2,
% \end{equation}
% \noindent
% where $y_i$ is the target result, $a_i$ and $s_i$ are the current action and state respectively, $a'_i$ and $s'_i$ are the next action and state respectively, and $r(s_i,a_i)$ is the reward for taking action $a_i$ at state $s_i$. On top of that, the policy is basically defined as follow:
% \begin{equation}
%     \pi_\theta(a_t|s_t) =
%     \begin{cases}
%     1 & \text{if } a_t =\argmaxA_{a_t}Q_\phi(s'_t,a'_t)\\
%     0              & \text{otherwise}
% \end{cases}
% \end{equation}
%Unlike other standard policy gradient methods that update the gradient per data sample,
\textbf{PPO} is a variant of PG that enables multiple epochs of minibatch updates to improve the sample complexity.
Vanilla PG performs one gradient update per data sample while PPO uses a novel surrogate objective function to enable multiple epochs of minibatch updates. It alternates between sampling data through interaction with the environment and optimizing the surrogate objective function using stochastic gradient ascent.
It performs updates that maximizes the reward function while ensuring the deviation from the previous policy is small by using a surrogate objective function. The loss function of PPO is defined as:%Equation~\ref{eq:ppo}.
\begin{equation}\label{eq:ppo}
%\vspace{-1cm}
L^{CLIP}(\theta) =
\hat{E}_{t}[ min( r_t(\theta)\hat{A}_t, clip(r_t(\theta), 1 - \varepsilon, 1 + \varepsilon) \hat{A}_t )]
\end{equation}
where $r_t(\theta)$ is defined as a probability ratio 
$\frac{\pi_\theta(\mathbf{a_t} | \mathbf{s_t})}{\pi_{\theta_{old}}(\mathbf{a_t} | \mathbf{s_t})}$
so $r(\theta_{old}) = 1$. This term penalizes policy update that move $r_t(\theta)$ from $r(\theta_{old})$.
$\hat{A}_t$ denotes the estimated advantage that approximates how good $\mathbf{a_t}$ is compared to the average. 
The second term in the $min$ function acts as a disincentive for moving $r_t$ outside of $[1-\varepsilon, 1+\varepsilon]$ where $\varepsilon$ is a hyperparameter.

\textbf{A3C} uses an actor (usually a neural network) that interacts with the critic, which is another network that evaluates the action by computing the value function. The critic tells the actor how good its action was and how it should adjust. The update performed by the algorithm
can be seen as $\nabla_\theta log\pi_\theta(a_{i,t}|s_{i,t})\hat{A}_t$.

\subsection{Evolutionary Algorithms}
Evolutionary algorithms are another technique that can be used to search for the best compiler pass ordering.  
It contains a family of population-based meta-heuristic optimization algorithms inspired by natural selection. The main idea of these algorithms is to sample a population of solutions and use the good ones to direct the distribution of future generations. Two commonly used Evolutionary Algorithms are Genetic Algorithms (GA)~\cite{goldberg2006genetic} and Evolution Strategies (ES)~\cite{conti2018improving}.

\textbf{GA} generally requires a genetic representation of the search space where the solutions are coded as integer vectors.  The algorithm starts with a pool of candidates, then iteratively evolves the pool to include solutions with higher fitness by the three following strategies: selection, crossover, and mutation. Selection keeps a subset of solutions with the highest fitness values. These selected solutions act as parents for the next generation. Crossover merges pairs from the parent solutions to produce new offsprings. Mutation perturbs the offspring solutions with a low probability. The process repeats until a solution that reaches the goal fitness is found or after a certain number of generations. 

\textbf{ES} works similarly to GA. However, the solutions are coded as real numbers in ES. In addition, ES is self-adapting. The hyperparameters, such as the step size or the mutation probability, are different for different solutions. They are encoded in each solution,  so good settings get to the next generation with good solutions. Recent work~\cite{salimans2017evolution} has used ES to update policy weights for RL and showed it is a good alternative for gradient-based methods. 

\section{AutoPhase Framework for Automatic Phase Ordering} 
\label{sec:framework}

\begin{figure}[t!]
    \centering
    \includegraphics[width=0.48\textwidth]{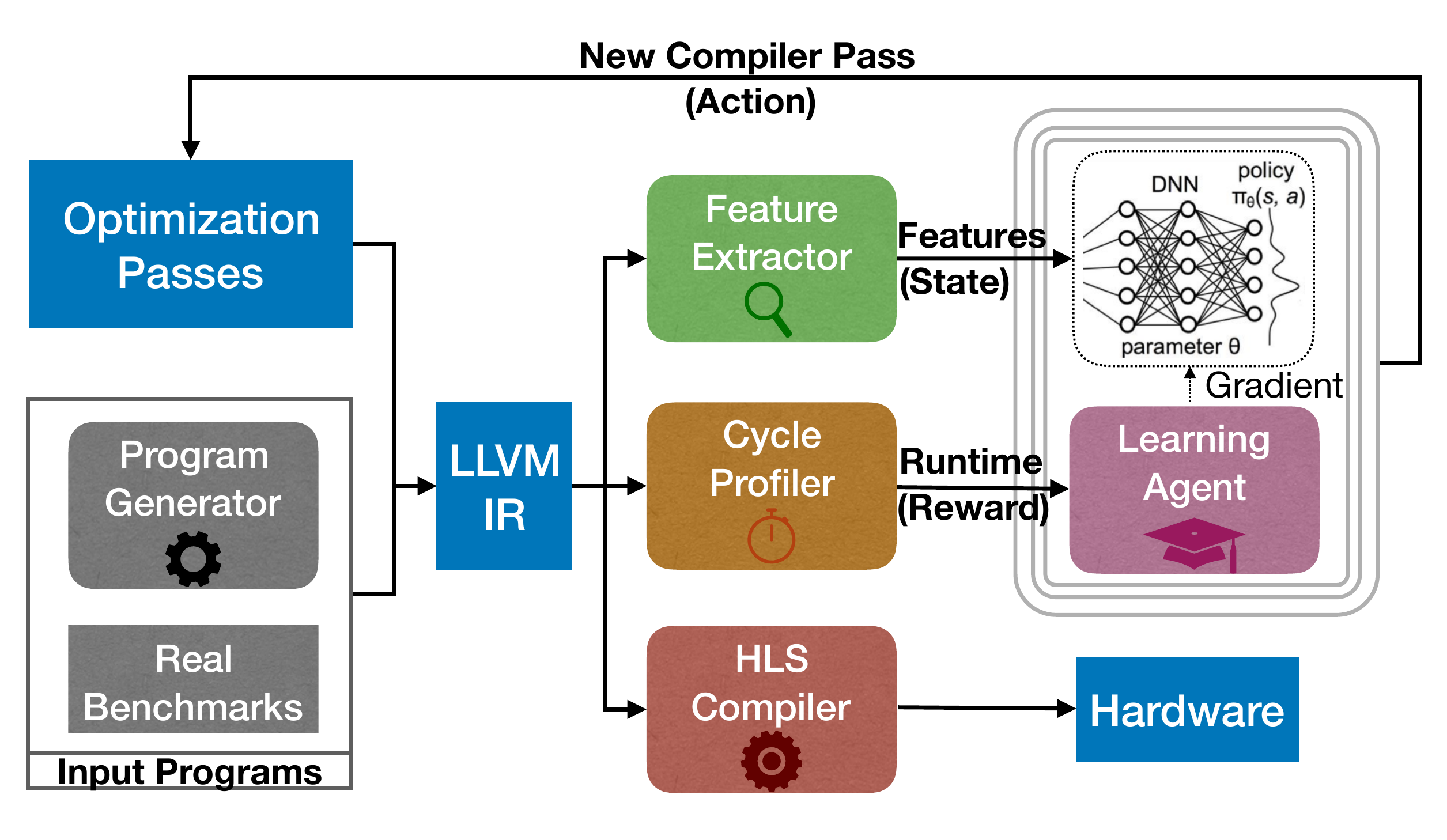}
    \caption{The block diagram of AutoPhase. The input programs are compiled to an LLVM IR using Clang/LLVM. The feature extractor and clock-cycle profiler are used to generate the input features (state) and the runtime improvement (reward), respectively from the IR. The input features and runtime improvement are fed to the deep RL agent as in input data to train on. The RL agent predicts the next best optimization passes to apply. After convergence, the HLS compiler is used to compile the LLVM IR to hardware RTL.}
    \label{fig:framework}
    \vspace{-0.5cm}
\end{figure}

We leverage an existing open-source HLS framework called LegUp~\cite{canis2013legup} that compiles a C program into a hardware RTL design. 
In \cite{huang2013effect}, an approach is devised to quickly determine the number of hardware execution cycles without requiring time-consuming logic simulation.
We develop our RL simulator environment based on the existing harness provided by LegUp and validate our final results by going through the time-consuming logic simulation. AutoPhase takes a program (or multiple programs) and intelligently explores the space of possible passes to figure out an optimal pass sequence to apply. Table~\ref{tab:passes} lists all the passes used in AutoPhase. The workflow of AutoPhase is illustrated in Figure~\ref{fig:framework}.

\subsection{HLS Compiler}
AutoPhase takes a set of programs as input and compiles them 
%front-end of the LLVM compiler. LLVM represents programs 
to a hardware-independent intermediate representation (IR) using the Clang front-end of the LLVM compiler. Optimization and analysis passes act as transformations on the IR, taking a program as input and emitting a new IR as output. The HLS tool LegUp is invoked after the compiler optimization as a back-end pass, which transforms LLVM IR into hardware modules.
%The compute logic is turned into a hardware datapath and the control logic is turned into a hardware FSM respectively by the HLS tool. 
%Its SDC-based scheduler \cite{cong2006efficient} operates at the basic block level to exploit the instruction-level parallelism. 
\subsection{Clock-cycle Profiler}
Once the hardware RTL is generated, one could run a hardware simulation to gather the cycle count results of the synthesized circuit. This process is quite time-consuming, hindering RL and all other optimization approaches. Therefore, we approximate cycle count using the profiler in LegUp~\cite{huang2013effect}, which leverages the software traces and runs $20\times$ faster than hardware simulation. 
In LegUp, the frequency of the generated circuits is set as a compiler constraint that directs the HLS scheduling algorithm. In other words, HLS tool will always try to generate hardware that can run at a certain frequency. In our experiment setting, without loss of generality, we set the target frequency of all generated hardware to 200MHz. We experimented with lower frequencies too; the improvements were similar but the cycle counts the different algorithms achieved were better as more logic could be fitted in a single cycle. 
%The profiler first runs the software program to gather information on the number of times each basic block is executed, then multiplies it with the clock cycle time of the basic block determined by the HLS scheduler to produce the total clock cycle time of the circuit. This software approach is $20\times$ faster than hardware simulation, which significantly reduces the runtime bottleneck of the deep reinforcement learning algorithm. 

\subsection{IR Feature Extractor}
Wang \textit{et al.}~\cite{wang2018} proposed to convert a program into an observation by extracting all the features from the program. Similarly, in addition to the LegUp backend tools, we developed analysis passes to extract 56 static features from the program, such as the number of basic blocks, branches, and instructions of various types. 
%Each individual feature is insufficient to guide the machine learning algorithm.
We use these features as partially observable states for the RL learning and hope the neural network can capture the correlation of certain combinations of these features and certain optimizations. Table \ref{tab:tab1} lists all the features used.
%All the features we gather are listed in Table~\ref{tab:tab1} in the appendix. 
%The framework takes a program or multiple ones and compiles them using the LLVM compiler, which was modified to output the program features. The HLS toolchain is used to extract the number of cycles that are used as the cost function to the RL architecture to find the optimal passes. 

\subsection{Random Program Generator}
As a data-driven approach, RL generalizes better if we train the agent on more programs.
However, there are a limited number of open-source HLS examples online. Therefore, we expand our training set by automatically generating synthetic HLS benchmarks. We first generate standard C programs using CSmith~\cite{yang2011csmith}, a random C program generator, which is originally designed to generate test cases for finding compiler bugs. Then, we develop scripts to filter out programs that take more than five minutes to run on CPU or fail the HLS compilation. 

\subsection{Overall Flow of AutoPhase}
We integrate the compilation utilities into a simulation environment in Python with APIs similar to an OpenAI gym~\cite{brockman2016openai}.
%The programs are compiled , features and cycles are extracted from the IR and HLS toolchain respectively, and are afterwards fed to the RL network to learn the optimal passes. 
%Our environment takes an input program and returns the program feature vectors and the total number of clock cycles. We consider two types of features as the state for the RL: the current program IR and the sequence of passes that have been applied. 
The overall flow works as follows:
\begin{enumerate}
\itemsep 0em 
\item The input program is compiled into LLVM IR using the Clang/LLVM. 
\item The IR Feature Extractor is run to extract salient program features. 
\item LegUp compiles the LLVM IR into hardware RTL.
\item The Clock-cycle Profiler estimates a clock-cycle count for the generated circuit. 
\item The RL agent takes the program features or the histogram of previously applied passes and the improvement in clock-cycle count as input data to train on. 
%are fed into the RL agent to derive a good LLVM optimization sequence.
%In RL specifically, the clock-cycle time is used for reward calculation, and the program features are used as partially observable states for the RL agents. 
\item The RL agent predicts the next best optimization passes to apply. 
%In iterative methods (\textit{i.e.}, greedy algorithm, genetic algorithm, and RL), the algorithm then predicts the next best action to take. In RL, the action would be the next optimization to apply to the end of an existing sequence of passes. In a modified greedy algorithm, the next action could be the next best place to insert the optimization pass.
\item New LLVM IR is generated after the new optimization sequence is applied. 
\item The machine learning algorithm iterates through steps (2)--(7) until convergence.
\end{enumerate}
Note that AutoPhase uses the LLVM compiler and the passes used are listed in Table~\ref{tab:tab1}. However, adding support for any compiler or optimization passes in AutoPhase is very easy and straightforward. The action and state definitions must be specified again. 
\begin{table*}[!t]
\caption{LLVM Transform Passes.}
\scriptsize
\centering
\begin{tabular}{ccccccccccc}
\hline
0 & 1 & 2 & 3 & 4 & 5 & 6 & 7 & 8 & 9 & 10 \\
-correlated-propagation & -scalarrepl & -lowerinvoke & -strip & -strip-nondebug & -sccp & -globalopt & -gvn & -jump-threading & -globaldce & -loop-unswitch \\
\end{tabular}
\begin{tabular}{ccccccccccc}
\hline
11 & 12 & 13 & 14 & 15 & 16 & 17 & 18 & 19 & 20 & 21 \\
-scalarrepl-ssa & -loop-reduce & -break-crit-edges & -loop-deletion & -reassociate & -lcssa & -codegenprepare & -memcpyopt & -functionattrs & -loop-idiom & -lowerswitch \\
\end{tabular}
\begin{tabular}{cccccccccccc}
\hline
22 & 23 & 24 & 25 & 26 & 27 & 28 & 29 & 30 & 31 & 32 & 33 \\
-constmerge & -loop-rotate & -partial-inliner & -inline & -early-cse & -indvars & -adce & -loop-simplify & -instcombine & -simplifycfg & -dse & -loop-unroll \\
\end{tabular}
\begin{tabular}{cccccccccccc}
\hline
34 & 35 & 36 & 37 & 38 & 39 & 40 & 41 & 42 & 43 & 44 & 45 \\
-lower-expect & -tailcallelim & -licm & -sink & -mem2reg & -prune-eh & -functionattrs & -ipsccp & -deadargelim & -sroa & -loweratomic & -terminate \\
\hline
\end{tabular}
\label{tab:passes}

\end{table*}
\begin{table*}[!t]
\scriptsize
\centering
\caption{Program Features.}
\begin{tabular}{
|c|c|c|c|}
\hline 0&
Number of BB where total args for phi nodes \textgreater 5 & 28&  Number of And insts \\ \hline 1&
Number of BB where total args for phi nodes is {[}1,5{]} & 29&  Number of BB's with instructions between {[}15,500{]} \\ \hline 2&
Number of BB's with 1 predecessor & 30&  Number of BB's with less than 15 instructions \\ \hline 3&
Number of BB's with 1 predecessor and 1 successor & 31&  Number of BitCast insts \\ \hline 4&
Number of BB's with 1 predecessor and 2 successors & 32&  Number of Br insts \\ \hline 5&
Number of BB's with 1 successor & 33&  Number of Call insts \\ \hline 6&
Number of BB's with 2 predecessors & 34&  Number of GetElementPtr insts \\ \hline 7&
Number of BB's with 2 predecessors and 1 successor & 35&  Number of ICmp insts \\ \hline 8&
Number of BB's with 2 predecessors and successors & 36&  Number of LShr insts \\ \hline 9&
Number of BB's with 2 successors & 37&  Number of Load insts \\ \hline 10&
Number of BB's with \textgreater{}2 predecessors & 38&  Number of Mul insts \\ \hline 11&
Number of BB's with Phi node \# in range (0,3{]} & 39&  Number of Or insts \\ \hline 12&
Number of BB's with more than 3 Phi nodes & 40&  Number of PHI insts \\ \hline 13&
Number of BB's with no Phi nodes & 41&  Number of Ret insts \\ \hline 14&
Number of Phi-nodes at beginning of BB & 42&  Number of SExt insts \\ \hline 15&
Number of branches & 43&  Number of Select insts \\ \hline 16&
Number of calls that return an int & 44&  Number of Shl insts \\ \hline 17&
Number of critical edges & 45&  Number of Store insts \\ \hline 18&
Number of edges & 46&  Number of Sub insts \\ \hline 19&
Number of occurrences of 32-bit integer constants & 47&  Number of Trunc insts \\ \hline 20&
Number of occurrences of 64-bit integer constants & 48&  Number of Xor insts \\ \hline 21&
Number of occurrences of constant 0 & 49&  Number of ZExt insts \\ \hline 22&
Number of occurrences of constant 1 & 50&  Number of basic blocks \\ \hline 23&
Number of unconditional branches & 51&  Number of instructions (of all types) \\ \hline 24&
Number of Binary operations with a constant operand & 52&  Number of memory instructions \\ \hline 25&
Number of AShr insts & 53&  Number of non-external functions \\ \hline 26&
Number of Add insts & 54&  Total arguments to Phi nodes \\ \hline 27&
Number of Alloca insts & 55&  Number of Unary operations \\ \hline
\end{tabular}
\label{tab:tab1}
\vspace{-0.1cm}
\end{table*}

\section{Correlation of Passes and Program Features} 

\label{sec:features}
%Among the major challenges that occur when applying machine learning techniques to solve different challenging problems is the
Similar to the case with many deep learning approaches, explainability is one of the major challenges we face when applying deep RL to the phase-ordering challenge. To analyze and understand the correlation of passes and program features, we use random forests~\cite{breiman2001random} to learn the importance of different features. Random forest is an ensemble of multiple decision trees. 
%that are used to make decisions on data. 
The prediction made by each tree could be explained by tracing the decisions made at each node and calculating the importance of different features on making the decisions at each node. 
This helps us to identify the effective features and passes to use and show whether our algorithms learn informative patterns on data.

For each pass, we build two random forests to predict whether applying it would improve the circuit performance. The first forest takes the program features as inputs while the second takes a histogram of previously applied passes. To gather the training data for the forests, we run PPO with high exploration parameter on 100 randomly generated programs to generate feature--action--reward tuples. The algorithm assigns higher importance to the input features that affect the final prediction more. 
%We represent the importance of each feature in heat maps.
%We analyze the heat map of the importance of each feature on making the final decision on whether the pass was helpful or not. In other words, if the feature was very helpful in classifying whether the pass will be helpful or not, the importance of this feature will be higher.

\subsection{Importance of Program Features}
\begin{figure}[!t]
    \centering
    \includegraphics[trim={9cm 3.5cm 4.2cm 0.8cm},clip,width=0.5\textwidth]{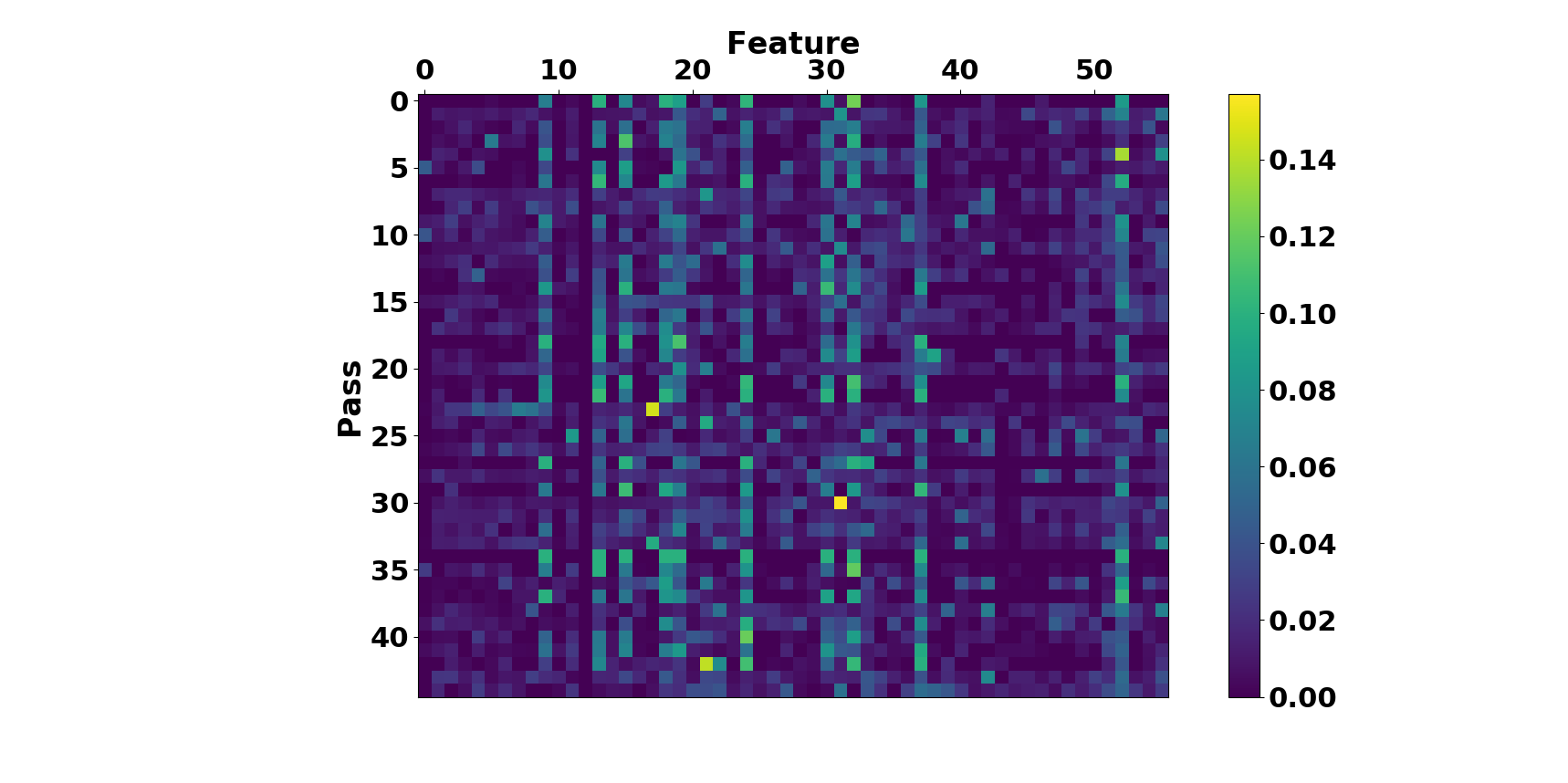}
    \caption{Heat map illustrating the importance of feature and pass indices.}
    \label{fig:heatmap1}
    \vspace{-0.1cm}
\end{figure}

The heat map in Figure~\ref{fig:heatmap1} shows the importance of different features on whether a pass should be applied. The higher the value is, the more important the feature is (the sum of the values in each row is one).
The random forest is trained with 150,000 samples generated from the random programs. The index mapping of features and passes can be found in Tables~\ref{tab:passes} and \ref{tab:tab1}. For example, the yellow pixel corresponding to feature index 17 and pass index 23 reflects
that \textit{number-of-critical-edges} affects the decision on whether to apply \textit{-loop-rotate} greatly.  A critical edge in control flow graph is an edge that is neither the only edge leaving its source block, nor the only edge entering its destination block. The critical edges can be commonly seen in a loop as a back edge so the number of critical edges might roughly represent the number of loops in a program. The transform pass \textit{-loop-rotate} detects a loop and
transforms a while loop to a do-while loop to eliminate one branch instruction in the loop body. Applying the pass results in better circuit performance as it reduces the total number of FSM states in a loop.  
%that applying pass \textit{-loop-rotate} was very impactful on feature \textit{number-of-critical-edges} since \textit{-loop-rotate} performs a simple loop rotation that changes the branch orderings resulting in different number of critical edges. In general this transformation was very helpful. 

Other expected behaviors are also observed in this figure. For instance, the correlation between \textit{number of branches} and the transform passes \textit{-loop-simplify}, \textit{-tailcallelism} (which transforms calls of the current function \textit{i.e.}, self recursion, followed by a return instruction with a branch to the entry of the function, creating a loop), \textit{-lowerswitch} (which rewrites switch instructions with a sequence of branches). Other interesting behaviors are also captured. For example, in the correlation between \textit{binary operations with a constant operand} and \textit{-functionattrs}, which marks different operands of a function as read-only (constant). 
Some correlations are harder to explain, for example, \textit{number of BitCast instructions} and \textit{-instcombine}, which combines  instructions into fewer simpler instructions. This is actually a result of \textit{-instcombine} reducing the loads and stores that call bitcast instructions for casting pointer types.  
%Apparently, in some cases, \textit{-instcombine} checks for the types and turns instructions with undefined behavior into unreachable, which directly lowers the number of BitCast instructions. 
Another example is \textit{number of memory instructions} and \textit{-sink}, where \textit{-sink} basically moves memory instructions into successor blocks and delays the execution of memory until needed.  Intuitively, whether to apply \textit{-sink} should be dependent on whether there is any memory instruction in the program. 
Our last example to show is \textit{number of occurrences of constant 0} and \textit{-deadargelim}, where \textit{-deadargelim} helped eliminate dead/unused constant zero arguments.

Overall, we observe that all the passes % except for pass number 26 (\textit{-early-cse}) 
are correlated to some features and are able to affect the final circuit performance. 
%a transformation of the program in a way that could improve or worsen its performance. %Pass \textit{-early-cse} was not very helpful because only a small fraction of the programs tested had trivially redundant computations that could be eliminated. Note that this was not the case for CHstone; in CHstone all the passes were useful, because CHstone includes a wide range of applications in different domains.
We also observe that multiple features are not effective at directing decisions and training with them could increase the variance that would result in lower prediction accuracy of our results. 
For example, the total number of instructions did not give a direct indication of whether applying a pass would be helpful or not. This is because sometimes more instructions could improve the performance (for example, due to loop unrolling) and eliminating unnecessary code could also improve the performance. In addition, the importance of features varies among different benchmarks depending on the tasks they perform.
\vspace{-0.2cm}
\subsection{Importance of Previously Applied Passes}
\begin{figure}[!t]
    \centering
    \includegraphics[trim={15cm 3.5cm 4.6cm 0.8cm},clip,width=0.5\textwidth]{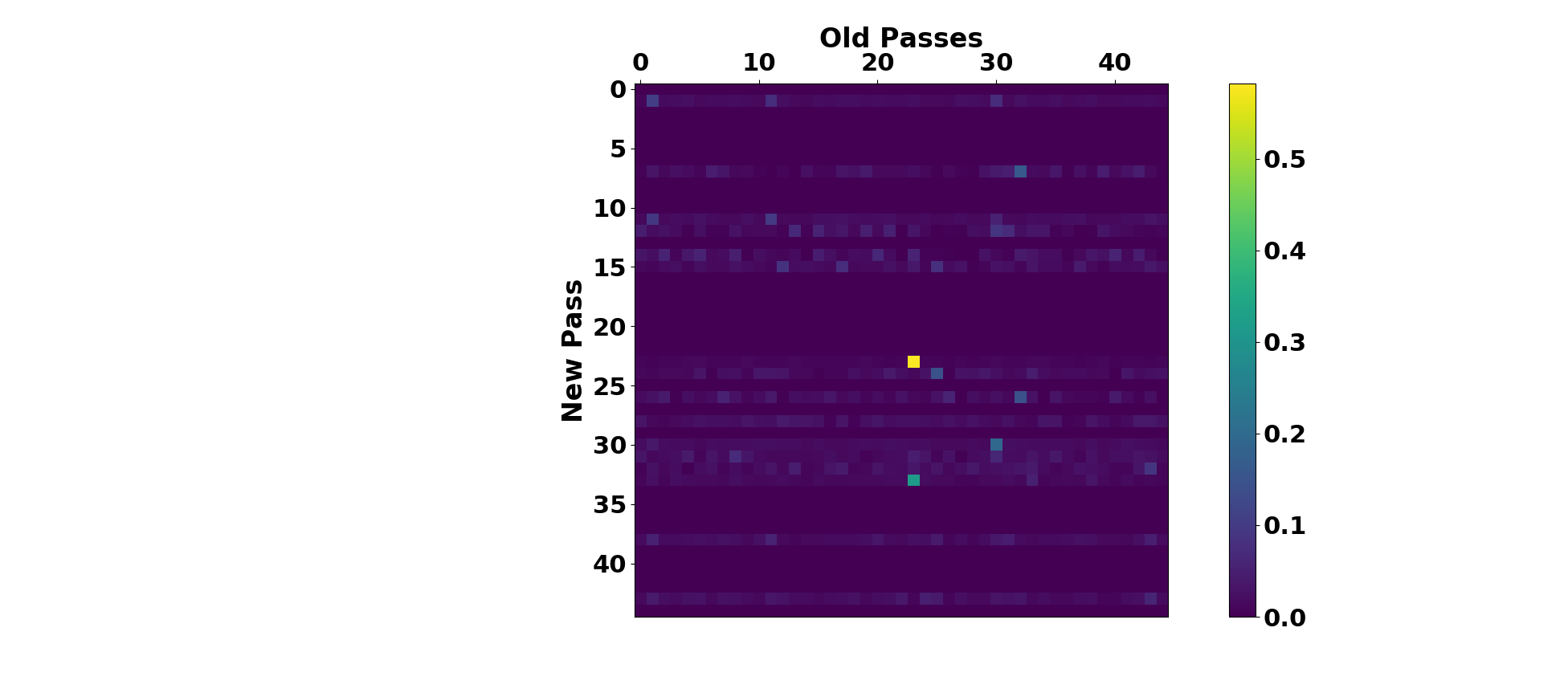}
    \caption{Heat map illustrating the importance of indices of previously applied passes and the new pass to apply.}
    \label{fig:heatmap2}
\end{figure}

Figure~\ref{fig:heatmap2} illustrates the impact of previously applied passes on the new pass to apply. The higher the value is, the more important having the old pass is. From this figure, we learn that for the programs we trained on passes \textit{-scalarrepl, -gvn, -scalarrepl-ssa, -loop-reduce, -loop-deletion, -reassociate, -loop-rotate, -partial-inliner, -early-cse, -adce, -instcombine, -simplifycfg, -dse, -loop-unroll, -mem2reg, and -sroa}, are more impactful on the performance compared to the rest of the passes regardless of their order in the trajectory. Point (23,23) has the highest importance in which implies that pass \textit{-loop-rotate} is very helpful and should be included if not applied before. By examining thousands of the programs, we find that \textit{-loop-rotate} indeed reduces the cycle count significantly. Interestingly, applying this pass twice is not harmful if the passes were given consecutively. However, giving this pass twice with some other passes between them is sometimes very harmful. Another interesting behavior our heat map captured is the fact that applying pass 33 (\textit{-loop-unroll}) after (not necessarily consecutive) pass 23 (\textit{-loop-rotate}) was much more useful compared to applying these two passes in the opposite order. 
%\section{Methodology} % 1 page 
\section{Problem Formulation}
\label{sec:DRLA}
\subsection{The RL Environment Definition}
\label{subsubsec:conf1}
Assume the optimal number of passes to apply is $N$ and there are $K$ transform passes to select from in total, our search space $\mathcal{S}$ for the phase-ordering problem is $[0, K^N)$. 
Given $M$ program features and the history of already applied passes, the goal of deep RL is to learn the next best optimization pass $a$ to apply that minimizes the long term cycle count of the generated hardware circuit. 
Note that the optimization state $s$ is partially observable in this case as the $M$ program features cannot fully capture all the properties of a program.  

\textbf{Action Space} -- 
we define our action space $\mathcal{A}$ as
$\{a \in \mathbb{Z}: a \in [0, K)\}$ where $K$ is the total number of transform passes.% defined in Table \ref{tab:passes}. 

\textbf{Observation Space} -- 
two types of input features were considered in our evaluation: 
\textbf{\textcircled{1} program features}  $\mathbf{o_f} \in 
\mathbb{Z}^M$ listed in Table \ref{tab:tab1} and 
\textbf{\textcircled{2} action history} which is a histogram of previously applied passes $\mathbf{o_a} \in \mathbb{Z}^K$. After each RL step where the pass $i$ is applied, we call the feature extractor in our environment to return new $\mathbf{o_f}$, and update the action histogram element $o_{a_i}$ to $o_{a_i} + 1$.  

\textbf{Reward} -- 
the cycle count of the generated circuit is reported by the clock-cycle profiler at each RL iteration. 
Our reward is defined as $R = c_{prev} - c_{cur}$, where $c_{prev}$ and $c_{cur}$ represent the previous and the current cycle count of the generated circuit respectively. It is possible to define a different reward for different objectives. For example, the reward could be defined as the negative of the area and thus the RL agent will optimize for the area. It is also possible to co-optimize multiple objectives (e.g., area, execution time, power, etc.) by defining a combination of different objectives.

\subsection{Applying Multiple Passes per Action}
\label{subsubsec:conf2}
An alternative to the action formulation above is to evaluate a complete sequence of passes with length $N$ instead of a single action $a$ at each RL iteration. Upon the start of training a new episode, the RL agent resets all pass indices $\mathbf{p} \in \mathbb{Z}^N$ to the index value $\frac{K}{2}$. For pass ${p_i}$ at index $i$, 
the next action to take is either to change to a new pass or not. 
By allowing positive and negative index update for each $p$, we reduced the total steps required to traverse all possible pass indices.  
The sub-action space $a_i$ for each pass is thus defined as $[-1, 0, 1]$.
The total action space $\mathcal{A}$ is defined as $[-1, 0, 1]^N$.
At each step, the RL agent predicts the updates $[a_1, a_2, ..., a_N]$ to N passes, and the current optimization sequence $[p_1, p_2, ..., p_N]$ is updated to $[p_1+a_1, p_2+a_2, ..., p_N+a_N]$.  %The observation space and reward of this formulation remain the same as in \mbox{Configuration 1}~\ref{subsubsec:conf1}.  

\subsection{Normalization Techniques}
\label{subsection:norm}
In order for the trained RL agent to work on new programs, we need to properly normalize the program features and rewards so they represent a meaningful state among different programs. 
In this work, we experiment with two techniques: 
\textcircled{1} taking the logarithm of program features or rewards and,   
\textcircled{2} normalizing to a parameter from the original input program that roughly depicts the problem size. 
For technique \textcircled{1}, note that taking the logarithm of the program features not only reduces their magnitude, it also correlates them in a different manner in the neural network. Since, $w_1\log(o_{f_1}) + w_2\log(o_{f_2}) = log(o_{f_1}^{w_1}o_{f_2}^{w_2})$,
the neural network is learning to correlate the products of features instead of a linear combination of them.  
For technique \textcircled{2}, we normalize the program features to the total number of instructions in the input program ($\mathbf{o_{f\_norm}} =\frac{\mathbf{o_{f}}}{o_{f_{51}}}$), which is feature \#51 in Table \ref{tab:tab1}.
%We normalize the reward to the the cycle count of the original input program $c_{orig}$, so $R_{norm}=\frac{c_{prev}-c_{cur}}{c_{orig}}$. 

\section{Evaluation} % 2 pages 
\label{sec:results}
To run our deep RL algorithms we use RLlib~\cite{liang2017rllib}, an open-source library for reinforcement learning that offers both high scalability and a unified API for a variety of applications. RLlib is built on top of Ray~\cite{moritz2018ray}, a high-performance distributed execution framework targeted at large-scale machine learning and reinforcement learning applications. We ran the framework on a four-core Intel i7-4765T CPU%~\cite{Intel2017}
with a Tesla K20c GPU% ~\cite{Nvidia2012}
for training and inference. 

We set our frequency constraint in HLS to 200MHz and use the number of clock cycles reported by the HLS profiler as the circuit performance metric.
In~\cite{huang2013effect}, results showed a one-to-one correspondence between the clock cycle count and the actual hardware execution time under certain frequency constraint. Therefore, better clock cycle count will lead to better hardware performance.
\subsection{Performance}
To evaluate the effectiveness of various algorithms for tackling the phase-ordering problem, we run them on nine real HLS benchmarks and compare the results based on the final HLS circuit performance and the sample efficiency against state-of-the-art approaches for overcoming the phase ordering, which include random search, Greedy Algorithms~\cite{huang2013effect}, OpenTuner~\cite{ansel2014opentuner}, and Genetic Algorithms~\cite{DEAP_JMLR2012}. These benchmarks are adapted from CHStone~\cite{hara2008chstone} and LegUp examples. They are: \emph{adpcm}, \emph{aes}, \emph{blowfish}, \emph{dhrystone}, \emph{gsm}, \emph{matmul}, \emph{mpeg2}, \emph{qsort}, and  \emph{sha}.
For this evaluation, the input features/rewards were not normalized, the pass length was set to 45, and each algorithm was run on a per-program basis. Table~\ref{tab3} lists the action and observation spaces used in all the deep RL algorithms.
\begin{table*}[!t]
\scriptsize
\centering
\caption{The observation and action spaces used in the different deep RL algorithms.}
\label{tab3}
\begin{tabu}{|[1.2pt]c|[1.2pt]c|c| c|c|c|[1.2pt]}
\tabucline[1.2pt]{-}

& \textbf{RL-PPO1} & \textbf{RL-PPO2} & \textbf{RL-PPO3} & \textbf{RL-A3C} & \textbf{RL-ES} \\ \hline
\tabucline[1.2pt]{-}

\textbf{Deep RL Algorithm} & PPO & PPO & PPO & A3C & ES \\ \hline
\textbf{Observation Space} & Program Features & Action History &  Action History + Program Features & Program Features & Program Features \\ \hline
\textbf{Action Space} & Single-Action & Single-Action & Multiple-Action & Single-Action & Single-Action \\ \hline
\tabucline[1.2pt]{-}

\end{tabu}
\end{table*}
 
\begin{figure}[!t]
    \centering
    \includegraphics[width=0.5\textwidth]{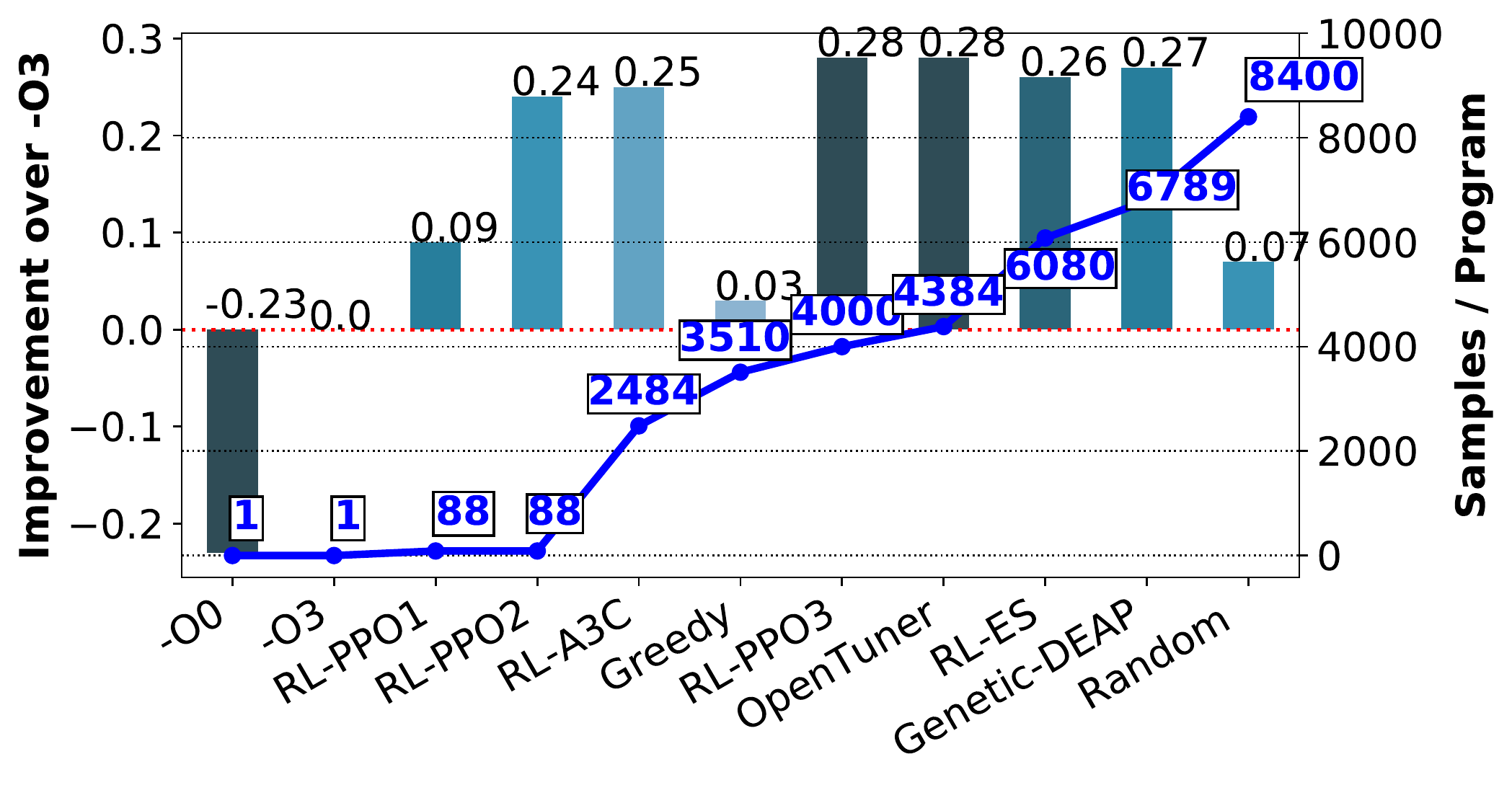}
    \vspace{-0.5cm}
    \caption{Circuit Speedup and Sample Size Comparison.}
    \label{fig:train_result}
\end{figure}

The bar chart in Figure~\ref{fig:train_result} shows the percentage improvement of the circuit performance compared to -O3 results on the nine real benchmarks from CHStone. The dots on the blue line in Figure~\ref{fig:train_result} show the total number of samples for each program, which is the number of times the algorithm calls the simulator to gather the cycle count.   
%Here is a summary on all the configurations we evaluate. 
\texttt{-O0} and \texttt{-O3} are the default compiler optimization levels. 
\texttt{RL-PPO1} is a PPO explorer where we set all the rewards to 0 to test if the rewards are meaningful. 
\texttt{RL-PPO2} is the PPO agent that learns the next pass based on a histogram of applied passes. 
\texttt{RL-A3C} is the A3C agent that learns based on the program features.
\texttt{Greedy} performs the greedy algorithm, which always inserts the pass that achieves the highest speedup at the best position (out of all possible positions it can be inserted to) in the current sequence. 
\texttt{RL-PPO3} uses a PPO agent and the program features but with the action space described in Section~\ref{subsubsec:conf2}.
explained in Section~\ref{subsubsec:conf2}. 
\texttt{OpenTuner} runs an ensemble of six algorithms, which includes two families of algorithms: particle swarm optimization~\cite{kennedy2010particle} and GA,
each with three different crossover settings. %PSO_GA_Bandit
\texttt{RL-ES} is similar to A3C agent that learns based on the program features, but updates the policy network using the evolution strategy instead of backpropagation.  
\texttt{Genetic-DEAP}~\cite{DEAP_JMLR2012} is a genetic algorithm implementation. 
\texttt{random} randomly generates a sequence of 45 passes at once instead of sampling them one-by-one.  

From \texttt{Greedy}, we see that always adding the pass in the current sequence that achieves the highest reward leads to sub-optimal circuit performance.
\texttt{RL-PPO2} achieves higher performance than \texttt{RL-PPO1}, which shows that the deep RL captures useful information during training. Using the histogram of applied passes results in better sample efficiency, but using the program features with more samples results in a slightly higher speedup.  
\texttt{RL-PPO2}, for example, at the minor cost of 4\% lower speedup, achieves 50$\times$ more sample efficiency than \texttt{OpenTuner}. 
Using ES to update the policy is supposed to be more sample efficient for problems with sparse rewards like ours, however, our experiments did not benefit from that. 
Furthermore, \texttt{RL-PPO3} with multiple action updates achieves a higher speedup than the other deep RL algorithms with a single action. One reason for that is  the ability of \texttt{RL-PPO3} to explore more passes per compilation as it applies multiple passes simultaneously in between every compilation. On the other hand, the other deep RL algorithms apply a single pass at a time. 

%OpenTuner and uses a bandit algorithm to pick the best algorithms to run. Move to background

%\vspace{-20pt}

\subsection{Generalization}
%Training on different programs and benchmarks and then using the trained network to inference on any existing program and expect it to achieve optimal results is not practical. Yet, 

With deep RL, the search should benefit from prior knowledge learned from other different programs. This knowledge should be transferable from one program to another. For example, as discussed in section~\ref{sec:features} applying pass \textit{-loop-rotate} is always beneficial, and \textit{-loop-unroll} should be applied after \textit{-loop-rotate}. Note that the black-box search algorithms, such as OpenTuner, GA, and greedy algorithms, cannot generalize. For these algorithms, rerunning a new search with many compilations is necessary for every new program, as they do not learn any patterns from the programs to direct the search and can be viewed as a smart random search. %Therefore, these algorithms need to perform a completely new search for every program. 

To evaluate how generalizable deep RL could be with different programs and whether any prior knowledge could be useful, we train on 100 randomly-generated programs using PPO. Random programs are used for transfer learning due to lack of sufficient benchmarks and because it is the worst-case scenario, \textit{i.e.}, they are very different from the programs that we use for inference. The improvement can be higher if we train on programs that are similar to the ones we inference on. We train a network with $256\times256$ fully connected layers and use the histogram of previously applied passes concatenated to the program features as the observation and passes as actions. 

%We normalized the program features to the total number of instructions.
As described in Section~\ref{subsection:norm}, we experiment with two normalization techniques for the program features: \textcircled{1} taking the logarithm of all the program features and \textcircled{2} normalizing the program features to the total number of instructions in the program.  
In each pass sequence, the intermediate reward was defined as the logarithm of the improvement in cycle count after applying each pass. The logarithm was chosen so that the RL agent will not give much larger weights to big rewards from programs with longer execution time. Three approaches were evaluated: \texttt{filtered-norm1}
uses the filtered (based on the analysis in Section~\ref{sec:features} where we only keep the important features and passes) program features and passes from Section with normalization technique \textcircled{1}, \texttt{original-norm2} uses all the program features and passes with normalization technique \textcircled{2}, and \texttt{filtered-norm2} uses the filtered program features and passes from Section~\ref{sec:features} with normalization technique \textcircled{2}. 
%data we analyzed from section~\ref{sec:features} where we keep only the impactful features and passes. 
Filtering the features and passes might not be ideal, especially when different programs have different feature characteristics and impactful passes. However, reducing the number of features and passes helps to reduce variance among all programs and significantly narrow the search space. % that will make the learning process easier and faster.

\begin{figure}[!t]
    \centering
    \includegraphics[width=\linewidth]{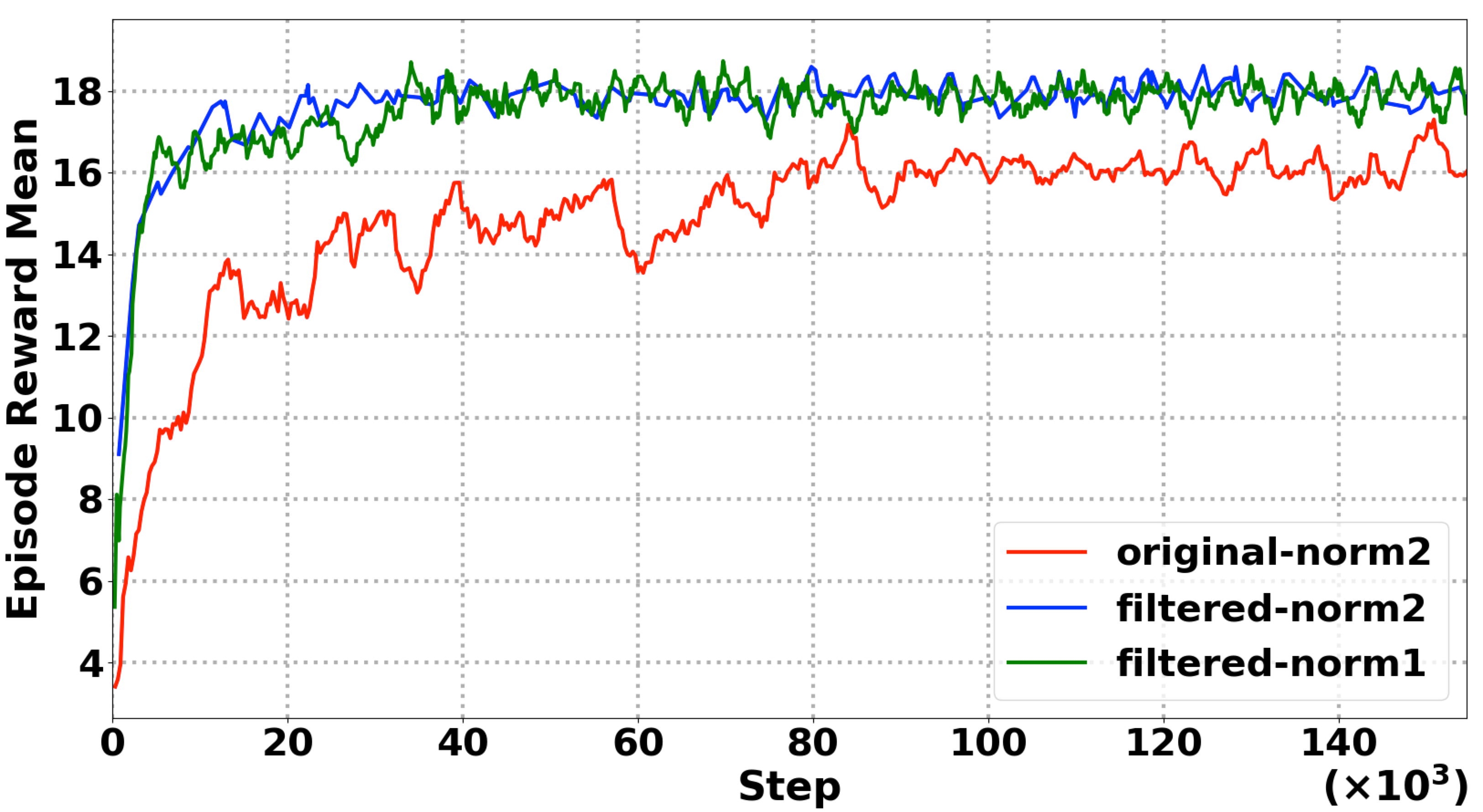}
    \caption{Episode reward mean as a function of step for the original approach where we use all the program features and passes and for the filtered approach where we filter the passes and features (with different normalization techniques). Higher values indicate faster circuit speed.}
    \label{fig:reward}
\end{figure}
Figure~\ref{fig:reward} shows the episode reward mean as a function of the step for the three approaches. We observe that \texttt{filtered-norm2} and \texttt{filtered-norm1} converge much faster and achieve a higher episode reward mean than \texttt{original-norm2}, which uses all the features and passes. At roughly 8,000 steps the \texttt{filtered-norm2} and \texttt{filter-norm1} already achieve a very high episode reward mean, with minor improvements in later steps. Furthermore, the episode reward mean of the filtered approaches is still higher than that of \texttt{original-norm2} even when we allowed it to train for 20 times more steps (\textit{i.e.}, 160,000 steps). This indicates that filtering the features and passes significantly improved the learning process. 
All three approaches learned to always apply pass \textit{-loop-rotate}, and \textit{-loop-unroll} after \textit{-loop-rotate}. Another useful pass that the three approaches learned to apply is \textit{-loop-simplify}, which performs several transformations to transform natural loops into a simpler form that enables subsequent analyses and transformations.

%We later use both networks to inference (rollout) the CHStone benchmarks and LegUp examples as well as other 12874 randomly generated programs. Interestingly, the network with all the features and passes delivered 7\% worse results than \mbox{-O3} while the filtered one delivered 6\% better results than -O3 with only 5\% of the programs performing slightly worse (less than 1\%) than -O3. While in general having more passes and features should give better performance, filtering the features and passes helped reduce the variance and enabled better learning of data that could be transferred to other programs more efficiently.
% simpler and more effective. 
%This pass will clean up blocks which are split out, but end up being unnecessary, so usage of this pass should not deteriorate generated code. 

\begin{figure}[!t]
    \centering
    \includegraphics[width=1\linewidth]{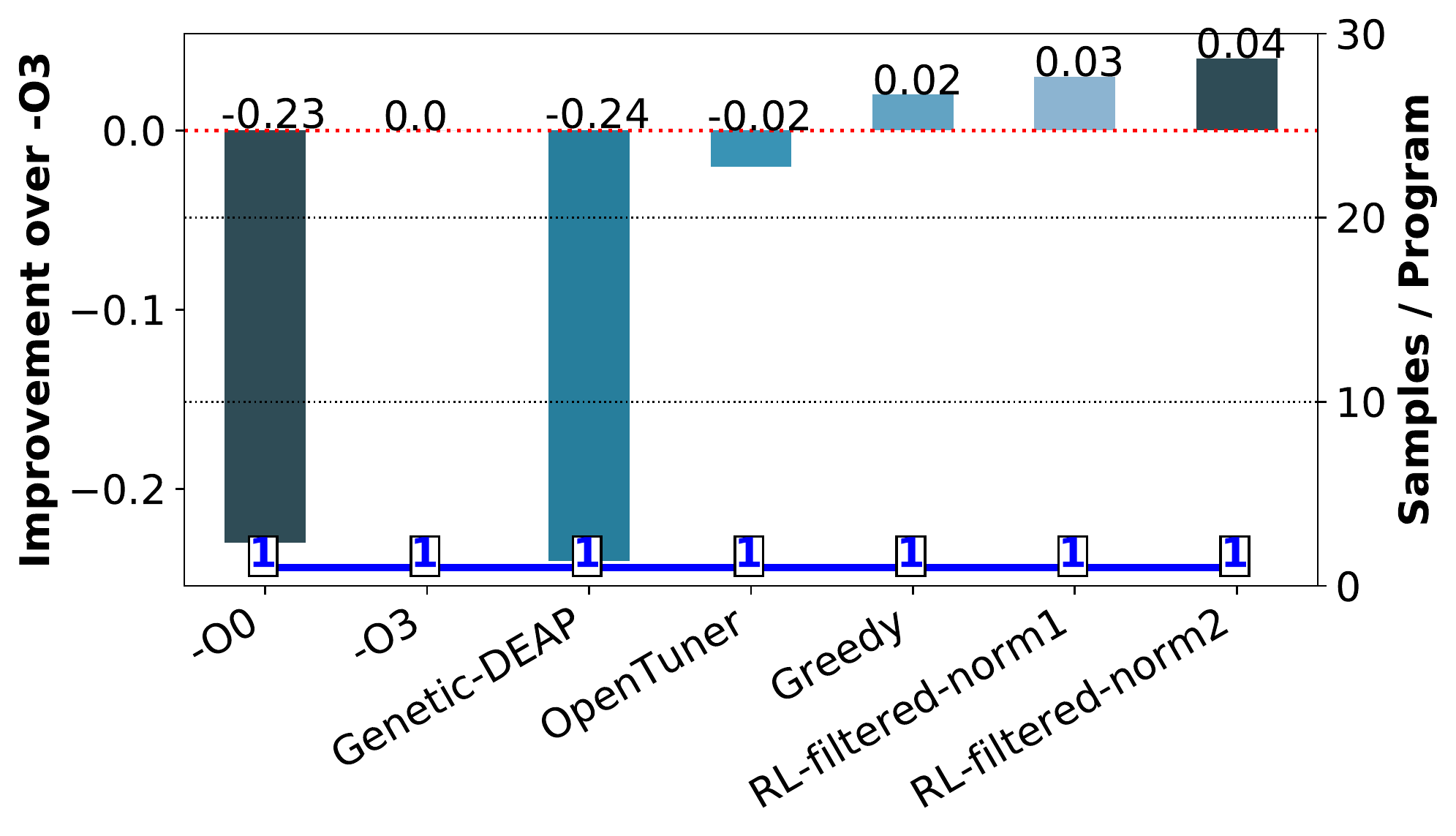}
    \vspace{-0.5cm}
    \caption{Circuit Speedup and Sample Size Comparison for deep RL Generalization.}
    \label{fig:train_result_generalization}
\end{figure}

We now compare the generalization results of \texttt{filtered-norm2} and \texttt{filtered-norm1} with the other black-box algorithms. We use 100 randomly-generated programs as the training set and nine real benchmarks from CHStone as the testing set for the deep RL-based methods.
With the state-of-the-art black-box algorithms, we first search for the best pass sequences that achieved the lowest aggregated hardware cycle counts for the 100 random programs and then directly apply them to the nine test set programs.   
In Figure~\ref{fig:train_result_generalization}, the bar chart shows the percentage improvement of the circuit performance compared to -O3 on the nine real benchmarks, the dots on the blue line show the total number of samples each inference takes for one new program.

This evaluation shows that the deep RL-based inference achieves higher speedup than the predetermined sequences produced by the state-of-the-art black-box algorithms for new programs. The predetermined sequences that are overfitted to the random programs can cause poor performance in unseen programs (\textit{e.g.,} -24\% for \texttt{Genetic-DEAP}).  
Besides, normalization technique \textcircled{2} works better compared to normalization technique\textcircled{1} for deep RL generalization (4\% vs 3\% speedup).
This indicates that normalizing the different instructions to the total number of instructions \textit{i.e.}, the distribution of the different instructions in Technique~\textcircled{2} represents more universal characteristics across different programs, 
while taking the log in Technique~\textcircled{1} only suppresses the value ranges of different program features. Furthermore, when we use other 12,874 randomly generated programs as the testing set with \texttt{filtered-norm2}, the speedup is 6\% compared to -O3.
%Lastly, the sample size of the RL methods is higher than one and is equal to the total number of actions to take in each RL trajectory for each inference. We set the trajectory length to 16 in this experiment.

\section{Conclusions} % 0.25 page 
\label{sec:conc}
\vspace{-0.1cm}
% In this work, we demonstrated a novel deep reinforcement learning based approach to robustly and intelligently improve the performance of HLS designs, by optimizing the compiler phase ordering. 
% These performance improvements require only a few minutes of training---one to two orders of magnitude faster than state-of-the-art approaches. 
% \JENNY{Add some numbers, how much better?}
% The techniques can be applied to software programs. 
% %The flexibility of our framework enables it to support optimizing any compiled program without being restricted to HLS. 
% We envision using such a framework to optimize a wide range of programs. 

In this paper, we propose an approach based on deep RL to improve the performance of HLS designs by optimizing the order in which the compiler applies optimization phases. We use random forests to analyze the relationship between program features and optimization passes. We then leverage this relationship to reduce the search space by identifying the most likely optimization phases to improve the performance, given the program features. Our RL based approach achieves 28\% better performance than compiling with the -O3 flag after training for a few minutes, and a 24\% improvement after training for less than a minute. Furthermore, we show that unlike prior work, our solution shows potential to generalize to a variety of programs.  
While in this paper we have applied deep RL to HLS, we believe that the same approach can be successfully applied to software compilation and optimization. Going forward, we envision using deep RL techniques to optimize a wide range of programs and systems.

%In this work, we show the potential of using RL to achieve a better ordering of compiler optimization passes. We built a framework that takes multiple programs and intelligently and robustly finds an optimal sequence of passes to apply. 
%We also show that using the program features solely is insufficient for RL to learn, due to limited observations, the necessity to apply multiple passes sometimes to affect such features, and inability to operate on multiple programs simultaneously.
%We propose to use program features or the applied passes as observations. 
%Significant performance and runtime benefits are achieved by using the later approach.
%allowing a novel approach based on RL to tackle the compiler phase ordering challenge and opening new horizons to explore in RL where actions could be used as observations. %More work should be done to better represent the programs.

\section*{Acknowledgement}
This research is supported in part by NSF CISE Expeditions Award CCF-1730628, the Defense Advanced Research Projects Agency (DARPA) through the Circuit Realization at Faster Timescales (CRAFT) Program under Grant HR0011-16-C0052, the Computing On Network Infrastructure for Pervasive Perception, Cognition and Action (CONIX) Research Center, NSF Grant 1533644, LANL Grant 531711, and DOE Grant DE-SC0019323, and gifts from Alibaba, Amazon Web Services, Ant Financial, CapitalOne, Ericsson, Facebook, Futurewei, Google, IBM, Intel, Microsoft, Nvidia, Scotiabank, Splunk, VMware, and ADEPT Lab industrial sponsors and affiliates. The views and opinions of authors expressed herein do not necessarily state or reflect those of the United States Government or any agency thereof.

\bibliographystyle{mlsys2020}
\bibliography{projBib}

\appendices
\begin{table*}[!t]
\centering
\caption{Program features: number of different operations.}
\begin{tabular}{|c|c|}
\hline
Number of BB where total args for phi nodes \textgreater 5 & Number of And insts \\ \hline
Number of BB where total args for phi nodes is {[}1,5{]} & Number of BB's with instructions between {[}15,500{]} \\ \hline
Number of BB's with 1 predecessor & Number of BB's with less than 15 instructions \\ \hline
Number of BB's with 1 predecessor and 1 successor & Number of BitCast insts \\ \hline
Number of BB's with 1 predecessor and 2 successors & Number of Br insts \\ \hline
Number of BB's with 1 successor & Number of Call insts \\ \hline
Number of BB's with 2 predecessors & Number of GetElementPtr insts \\ \hline
Number of BB's with 2 predecessors and 1 successor & Number of ICmp insts \\ \hline
Number of BB's with 2 predecessors and successors & Number of LShr insts \\ \hline
Number of BB's with 2 successors & Number of Load insts \\ \hline
Number of BB's with \textgreater{}2 predecessors & Number of Mul insts \\ \hline
Number of BB's with Phi node \# in range (0,3{]} & Number of Or insts \\ \hline
Number of BB's with more than 3 Phi nodes & Number of PHI insts \\ \hline
Number of BB's with no Phi nodes & Number of Ret insts \\ \hline
Number of Phi-nodes at beginning of BB & Number of SExt insts \\ \hline
Number of branches & Number of Select insts \\ \hline
Number of calls that return an int & Number of Shl insts \\ \hline
Number of critical edges & Number of Store insts \\ \hline
Number of edges & Number of Sub insts \\ \hline
Number of occurrences of 32-bit integer constants & Number of Trunc insts \\ \hline
Number of occurrences of 64-bit integer constants & Number of Xor insts \\ \hline
Number of occurrences of constant 0 & Number of ZExt insts \\ \hline
Number of occurrences of constant 1 & Number of basic blocks \\ \hline
Number of unconditional branches & Number of instructions (of all types) \\ \hline
Number of Binary operations with a constant operand & Number of memory instructions \\ \hline
Number of AShr insts & Number of non-external functions \\ \hline
Number of Add insts & Total arguments to Phi nodes \\ \hline
Number of Alloca insts & Number of Unary operations \\ \hline
\end{tabular}
\label{tab:tab1}
\end{table*}

\begin{table*}[!t]
\caption{The cycles for searching the best three passes using the different search algorithms for different tasks.}
\label{tab:3pass}
\hskip3.3cm\begin{tabular}{|c|c|c|c|c|c|c|}
\hline
\textbf{\textbf{}}  & \multicolumn{6}{c|}{\textbf{Cycles}}                                                                                                                                                                \\ \hline
\textbf{Task}       & \textbf{RL}        & \textbf{-O3}      & \textbf{random}    & \textbf{Greedy}     & \textbf{Genetic} & \textbf{No Opt.} \\ \hline
\textbf{aes}        & \textcolor{blue}{9003}   & 9181                               & \textcolor{blue}{9003}   & 11765                                & \textcolor{blue}{9003}                              & 11913   \\ \hline
\textbf{adpcm}      & \textcolor{blue}{13707}  & 16244                              & 15837                               & Error                                   & \textcolor{blue}{13707}                               & 40430   \\ \hline
\textbf{bf}         & \textcolor{blue}{180050} & 187327                             & 185250                              & 185269                               & \textcolor{blue}{180050}                              & 197395  \\ \hline
\textbf{jpeg}       & 1296330                             & 1313851                            & 1318382                             & \textcolor{blue}{1239409} & 1296330                              & 1474020 \\ \hline
\textbf{mpeg2}      & 8356                                & \textcolor{blue}{8260}  & 8356                                & 8469                                 & 8356                              & 10485   \\ \hline
\textbf{sha driver} & \textcolor{blue}{209154} & 226234                             & \textcolor{blue}{209154} & 222024                               & \textcolor{blue}{209154}                               & 296387  \\ \hline
\textbf{gsm}        & \textcolor{blue}{6168}   & 6491                               & 6172                                & 6869                                 & \textcolor{blue}{6168}                               & 7810    \\ \hline
\textbf{fir}        & 1850                                & \textcolor{blue}{1768}  & 1850                                & 2598                                 & 1850                              & 2172    \\ \hline
\textbf{dhry}       & 7767                                & \textcolor{blue}{5912}  & 8793                                & 8793                                 & 7767                              & 9194    \\ \hline
\textbf{qsort}      & \textcolor{blue}{49948}  & \textcolor{blue}{49948} & 53733                               & 54347                                & \textcolor{blue}{49948}                              & 56092   \\ \hline
\textbf{mm}         & \textcolor{blue}{33244}  & \textcolor{blue}{33244} & \textcolor{blue}{33244}  & \textcolor{blue}{33244}   & \textcolor{blue}{33244}                              & 42085   \\ \hline
\textbf{dfadd}      & \textcolor{blue}{726}    & \textcolor{blue}{726}   & \textcolor{blue}{726}    & \textcolor{blue}{726}     & \textcolor{blue}{726}                              & 773     \\ \hline
\end{tabular}
\end{table*}

\begin{table*}[]
\caption{The cycles for searching the best $12$ passes using the different search algorithms for different tasks.}
\label{tab:12pass}
\hskip0.3cm\begin{tabular}{|c|c|c|c|c|c|c|c|c|c|}
\hline
\textbf{}                            & \multicolumn{9}{c|}{\textbf{Cycles}}                             
\\ \hline
\textbf{Task}       & \textbf{DQN 12-steps} & \textbf{DQN 4-steps} & \textbf{DQN 1-step} & \textbf{PG}   & \textbf{-O3}      & \textbf{random}   & \textbf{Greedy}    & \textbf{Genetic}   & \textbf{No Opt.} \\ \hline
\textbf{aes}        & \textcolor{blue}{9003}      & \textcolor{blue}{9003}     & 11693                                 & \textcolor{blue}{9003}  & 9181                               & \textcolor{blue}{9003}  & \textcolor{blue}{9003}   & \textcolor{blue}{9003}      & 11913                             \\ \hline
\textbf{adpcm}      & \textcolor{blue}{8501}      & 13707                                 & 36537                                 & 13657                           & 16244                              & 14463                              & 13607                               & \textcolor{blue}{8501}   & 40430                             \\ \hline
\textbf{bf}         & 180054                                 & \textcolor{blue}{180050}   & 180069                                & \textcolor{blue}{180050}                           & 187327                             & 180054                             & \textcolor{blue}{180050} & 180054                              & 197395                            \\ \hline
\textbf{jpeg}       & 1305864                                & 1305583  & \textcolor{blue}{1295373}  & 1296330                           & 1313851                            & 1324660                            & 1305583                             & 1305775                             & 1474020                           \\ \hline
\textbf{mpeg2}      & 8356                                   & 8356                                  & 8392                                  & 8280                           & \textcolor{blue}{8260}  & 8280                               & 8274                                & 8280                                & 10485                             \\ \hline
\textbf{sha driver} & \textcolor{blue}{209154}    & \textcolor{blue}{209154}   & 222024                                & \textcolor{blue}{209154}                          & 226234                             & 209168                             & \textcolor{blue}{209154} & \textcolor{blue}{209154} & 269387                            \\ \hline
\textbf{gsm}        & \textcolor{blue}{5849}      & \textcolor{blue}{5849}     & 6696                                  & 6168                           & 6491                               & 6168                               & \textcolor{blue}{5849}   & \textcolor{blue}{5849}   & 7810                              \\ \hline
\textbf{fir}        & 1850                                   & 1850                                  & 1850                                  & 1850                           & \textcolor{blue}{1768}  & 1850                               & 1850                                & 1850                                & 2172                              \\ \hline
\textbf{dhry}       & 6162                                   & 7955                                  & 7767                                  & 7767                           & \textcolor{blue}{5912}  & 5974                               & 7767                                & 5974                                & 9194                              \\ \hline
\textbf{qsort}      & \textcolor{blue}{49948}     & \textcolor{blue}{49948}    & 49889                                 & \textcolor{blue}{49948}                           & \textcolor{blue}{49948} & 49889                              & 49889                               & \textcolor{blue}{49948}  & 56092                             \\ \hline
\textbf{mm}         & \textcolor{blue}{33244}     & \textcolor{blue}{33244}    & 33244                                 & \textcolor{blue}{33244}                           & \textcolor{blue}{33244} & \textcolor{blue}{33244} & \textcolor{blue}{33244}  & \textcolor{blue}{33244}  & 42085                             \\ \hline
\textbf{dfadd}      & \textcolor{blue}{726}       & \textcolor{blue}{726}      & 726                                   & \textcolor{blue}{726}                           & \textcolor{blue}{726}   & \textcolor{blue}{726}   & \textcolor{blue}{726}    & \textcolor{blue}{726}    & 773                               \\ \hline
\end{tabular}
\end{table*}

\end{document}